\documentstyle[proceedings,numreferences]{crckapb} 

\begin{opening}
\title{PHONONLIKE EXCITATIONS OF INSTANTON LIQUID\protect\\
AND NEW SCALE OF NON-PERTURBATIVE QCD}
\author{S.V. MOLODTSOV}
\institute{State Research Center,\\
 Institute of Theoretical and Experimental Physics,\\
117259, Moscow, Russia}
\author{A.M. SNIGIREV}
\institute{Nuclear Physics Institute, Moscow State University,\\
119899, Moscow, Russia}
\author{G.M. ZINOVJEV}
\institute{Bogolyubov Institute for Theoretical Physics,\\
National Academy of Sciences of Ukraine,\\
25143, Kiev, Ukraine}
\institute{Fakult\"at f\"ur Physik, Universit\'at Bielefeld,\\
D-33501, Bielefeld, Germany}
\end{opening}

\runningtitle{PHONONLIKE EXCITATIONS}

\begin{document}

\begin{abstract}
The phononlike excitations of (anti-)instanton ($\bar II$) liquid due to
adiabatic variations of vacuum wave functions are studied in this
paper. The kinetic energy term is found and the proper effective 
Lagrangian for such excitations is evaluated. The properties of their 
spectrum, while corresponding masses are defined by $\Lambda_{QCD}$,
are investigated. 
\end{abstract}

The model of (anti-)instanton liquid correctly seizes many 
nonperturbative phenomena and important vacuum features such as 
chiral symmetry breaking, the presence of gluon condensate and topological 
susceptibility \cite{1},\cite{2}. 
It is usually supposed that the corresponding functional integral in this
approach is saturated by quasi-classical configurations close to the exact 
solutions of the Yang-Mills equations (the Euclidean solutions called the
(anti-)instantons) and the wave function of vacuum, being homogeneous in
metric space, is properly reproduced by averaging over their collective
coordinates. In the $\bar II$ liquid approach one takes the superposition 
ansatz of the pseudo-particle (PP) fields as one of the simplest relevant 
approximations to the 'genuine' vacuum configuration 
\begin{equation}
\label{1} 
A_\mu(x)=\sum_{i=1}^N A_\mu(x;\gamma_i)~.
\end{equation}
Here $A_\mu(x;\gamma_i)$ denotes the field of a singled (anti-)instanton
in singular gauge with $4N_c$ (for the $SU(N_c)$ group) coordinates  
$\gamma=(\rho,~z,~\Omega)$, of size $\rho$ with the coordinate of its 
centre $z$, $\Omega$ as its colour orientation and
\begin{equation}
\label{2}
A^{a}_\mu(x;\gamma)=\frac {2}{g}~\Omega^{ab} \bar \eta_{b\mu\nu}
\frac{y_\nu}{y^2}\frac{\rho^2}{y^2+\rho^2}~,~~y=x-z~,
\end{equation}
where $\eta$ is the 't Hooft symbol \cite{3}.
For anti-instanton $\bar\eta\to\eta$ (making the choice of the singular
gauge allows us to sum up the solution preserving the asymptotic
behaviour). For simplicity we shall not introduce different symbols
for instanton and anti-instanton, and then in the superposition of
Eq.(\ref{1}) $N$ implies the PP total number in the 4-volume $V$
system 
with the density $n=N/V$. 
The action of the instanton liquid model is introduced by the 
following functional
\begin{equation}
\label{3}
\langle S \rangle= \int d^4 z \int d\rho~ n(\rho) s(\rho)~.
\end{equation}
The integration should be performed over the system volume along 
with averaging the one instanton action $s(\rho)$ weighted by 
instanton size distribution function 
$n(\rho)$. Then an action per one instanton is given by well-known
expression
\begin{equation}
\label{4}
s_1(\rho)=\beta(\rho)+5 \ln(\Lambda\rho)-\ln \widetilde \beta^{2N_c}+
\beta\xi^2\rho^2\int d\rho_1 n(\rho_1) \rho_1^{2}~,
\end{equation}
with the Gell-Mann-Low beta function 
$
\beta(\rho)=-\ln C_{N_c}-b \ln(\Lambda \rho)~,~~
\Lambda=\Lambda_{\overline{MS}}=0.92 \Lambda_{P.V.}~,$ 
and constant $C_{N_c}$ depending on the regularization scheme, here
$C_{N_c}\approx\frac{4.66~\exp(-1.68 N_c)}{\pi^2(N_c-1)!(N_c-2)!},
~b=\frac{11}{3} N_c~,
$
and the parameters $\beta=\beta(\bar\rho)$ and $\widetilde \beta=\beta +
\ln C_{N_c}$ are 
the $\beta$ function values at the fixed quantity of $\bar\rho$ 
(average instanton size).

Some terms (inperfect $\rho$ dependence) of Eq.(\ref{4}) could be obtained 
in the classical Yang-Mills theory with one-loop (quantum) corrections taken 
into account and resulting in a modification of coupling constant $g$ at the 
distinct scales. Indeed, the first term is the one instanton action 
$8\pi^2/g^2$ with the $\rho$-dependence of $g$ corrected. The last term of 
Eq.(\ref{4}) describes the pair interaction of PP in the instanton
ensemble with the constant $\xi$ characterizing, in a sense, the intensity 
of interaction $\xi^2=\frac{27}{4}\frac{N_c}{N_c^{2}-1} \pi^2$. 
The smallness of characteristic instanton liquid parameter 
$n\rho^4$ ('packing fraction') allows us
to drop out the $\rho$ dependence of the $\beta$-function. 
The logarithmic terms correspond to the pre-exponential factor 
contribution to the 
functional integral and are of a genuine quantum nature.
In the second term the $\rho^5$ factor makes the integration elements 
$d\rho$ and $d^4z$ dimensionless. Finally, the third term is logarithm 
of a square root of the one instanton action $\widetilde \beta$
raised to the power $4 N_c$. The latter is just the zero mode number 
of the one instanton solution and the corresponding $\rho$ dependence may
be again omitted because of small logarithmic contribution.

Taking the exponential form for the distribution function over the instanton 
action $n(\rho)\sim e^{-s_1(\rho)}$, we obtain directly from Eq.(\ref{3}) 
the self-consistent description of equilibrium state of instanton liquid with 
the well-known ground state 
$\mu(\rho)= \rho^{-5} \widetilde\beta^{2 N_c}
e^{-\beta(\rho)-\nu \rho^2/\overline{\rho^2}},$
where \\
$\nu=(b-4)/2,~
\left(\overline{\rho^2}\right)^2=\frac{\nu}{\beta \xi^2 n},~
n=\int d\rho~n(\rho),~
\overline{\rho^2}=\int d\rho~\rho^2n(\rho)/n.$

This argument corresponds to the maximum principle of \cite{2}. 
Approaching the functional of the averaged 
action per one instanton as a local form 
$\langle s_1 \rangle= \int d\rho~ s_1(\rho) n(\rho)/n$ where
$s_1(\rho)=\beta(\rho)+5 \ln(\Lambda\rho)-\ln \widetilde \beta^{2N_c}+
\beta\xi^2\rho^2n\overline{\rho^2}$, 
using $n(\rho)=C e^{-s(\rho)}$ ($C=const$) as a distribution 
function 
(actually it makes the problem self-consistent because an ~equilibrium 
distribution function should be dependent on an action only) and taking
the variation of
$\langle s \rangle-\langle s_1 \rangle= \int d\rho~\{ s(\rho)- s_1(\rho)\}
e^{-s(\rho)}/n$ over $s(\rho)$ one may come to the result 
$s(\rho)=s_1(\rho)+const$ keeping into the mind an arbitrary
normalization. Then maximizing the action per one instanton over 
$\bar\rho$, for example, we obtain an explicit description (E.D.) of 
ground state of the instanton liquid 
through the equilibrium value of $\bar\rho=e^{-\frac{2N_c}{2\nu-1}},$
close to the values obtained in \cite{2} (D.P.) 
\begin{center}
\begin{tabular}{cccccccccc}
&&&D.P.&&&&&~E.D.~~&\\\hline
      &$N_f$&$\bar\rho\Lambda$&$n/\Lambda^4$  &$\beta$
&~~~~~~~~~~&$N_f$&$\bar\rho\Lambda$&$n/\Lambda^4$  &$\beta$\\\hline
&0 & 0.37  &0.44&17.48 &
&0 & 0.37  &0.44&17.48\\      
&1 & 0.30  &0.81&18.86 &
&1 & 0.33  &0.63&18.10\\
&2 & 0.24  &1.59&20.12 & 
&2 & 0.28  &1.03&18.93\\
&3 & 0.19  &3.45&21.34 &
&3 & 0.22  &2.02&19.99\\
\hline
\end{tabular}
\end{center}
where $N_f$ is the number of flavours, $N_c=3$.

The distribution $\mu(\rho)$ has obvious physical meaning, namely,
the quantity $d^4z~d \rho~\mu(\rho)$ is proportional to the probability to 
find an instanton of size $\rho$ at some point of a volume element $d^4 z$. 
At small $\rho$ the behaviour of distribution function is stipulated by the 
quantum-mechanical uncertainty principle preventing a solution being 
compressed at a point (radiative correction). At large $\rho$ the constraint 
comes from the repulsive interaction between the PP which is amplified with
(anti-)instanton size increasing.

Deriving Eq.(\ref{3}) we should average over the instanton positions 
in a metric space. It is clear that the characteristic size, which has to 
be taken into account, should be larger enough than the mean instanton size
$\bar\rho$. But it should not be too large because the far ranged fragments of 
instanton liquid are not 'causally' dependent. The vacuum wave function is
expected to be homogeneous on this 
scale $L\ge \bar R$ ($\bar R$ is an average distance between the PP).
Let us 
remind that each PP contributes to the functional integral with the weight 
factor proportional to $1/V,~ V=L^4$. The characteristic configuration 
saturating the functional integral
is taken as the superposition (\ref{1}) with $N$ pseudoparticles in the 
volume $V$.
If one supposes that the PP number in an ensemble is still appropriate to 
consider them 
separately, then denoting  $\triangle N(\rho_i)$ as the PP number of size 
$\rho\in(\rho_i,~\rho_i+\triangle\rho)$, $K$ as the number of 
partitions within the interval $(\rho_{i},~\rho_{f})$, Eq.(\ref{1}) may be 
rewritten 
in the following form
\begin{equation}
\label{5}
A_\mu(x)=\sum^{K}_{i=1}\sum^{\triangle N(\rho_i)}_{j=1}
A_\mu(x;i,\gamma_j)~,
\end{equation}
where $A_\mu(x;i,\gamma_j)$ is the (anti-)instanton solution with the
calibrated size 
and  
$\gamma=(z,~\Omega)$ stands for the coordinate of its centre and colour 
orientation.
By definition $\sum^{K}_{i=1}\triangle N(\rho_i)=N.$
Further, introducing the distribution function 
$
n(\rho)=\frac{\triangle N(\rho)}{\triangle \rho}\frac{1}{V},
$
and normalizing it as $\sum^{K}_{i=1}n(\rho_i)~\triangle\rho~V=N$
(in the continual limit  $\triangle \rho \to 0$ it is valid 
$V\int d\rho~n(\rho)=N$) 
one can calculate the classical action $S_c =\frac14 \int d^4x~ 
G^2_{\mu\nu}$ 
of this configuration
averaging over the instanton positions in the metric and colour spaces.
As a result (with the superposition ansatz Eq.(\ref{1})) one instanton 
actions and 
the PP pair interactions only contribute to the average system action
$$
\langle S_c \rangle=\prod^N_{i=1}\int \frac{d^4 z_i}{V}
d \Omega_i~S_c~= \int d^4 z \int d\rho~ n(\rho)
\left\{\frac{8\pi^2}{g^2}+\frac{8\pi^2}{g^2}\xi^2\rho^2\int d\rho_1
 n(\rho_1) \rho_1^{2}\right\}
$$
where $d\Omega$ is a measure in the colour space with the unit
normalization. 
As above mentioned $\langle S_c\rangle$ including one loop corrections 
will then contribute to the functional integral.

It is easy to understand that Eq.(\ref{3}) describes properly even 
non-equi\-lib\-rium
states of the instanton liquid when the distribution function $n(\rho)$ 
does not coincide with the ground state one $\mu(\rho)$. Moreover, 
it allows us to generalize Eq.(\ref{3}) 
for the non-homogeneous liquid, when the size of the non-homogeneity 
obeys the obvious constraint $\lambda \ge L> \bar\rho$.

In what follows, we study the excitations of $\bar II$ liquid induced
by adiabatic dilatational deformations of the instanton solutions. 
Then, as the configurations saturating the functional integral we consider 
not the instanton solution itself but the quasizero modes which are 
parametrically very close in the functional space (a direction does exist 
where the action varies slowly) to the zero modes. 
The guiding idea of selecting a deformation originates from transparent
observation. The deformations measured in units of the action 
$\frac{dq~dp}{2\pi\hbar}$ (here $q,~p$ are the generalized coordinate and
momentum) have a physical meaning only. However, the instantons are 
characterized by 'static' coordinates $\gamma$ and, therefore, need to
appoint the conjugated momenta. It looks quite natural for the variable,
for example, $\rho$ to introduce those as $\dot\rho=d \rho/dx_4$.

Let us calculate first of all the corrections for the one-instanton
action. Dealing with superposition ansatz Eq.(\ref{1}) again, one should
include additional contribution to the chromoelectric field 
 \begin{equation}
\label{7}
G'^a_{\mu\nu}=G^a_{\mu\nu}+g^a_{\mu\nu}~.
\end{equation}
with the first term of strength tensor (s.t.) corresponding to the
contribution
generated by the 
instanton profile 
$$ G^a_{\mu\nu}=-\frac{8}{g}\frac{\rho^2}{(y^2+\rho^2)^2} 
\left (\frac12 \bar\eta_{a\mu\nu} + \bar\eta_{a\nu\rho}
\frac{y_\mu y_\rho}{y^2}
-\bar\eta_{a\mu\rho}\frac{y_\nu y_\rho}{y^2}
\right)~,
$$
and in adiabatic approximation the corrections have the form
$$ g^a_{4i}=\frac{\partial A^a_i}{\partial\rho}\dot\rho=
\frac4g\bar\eta_{ai\nu}~\frac{y_\nu~\rho}{(y^2+\rho^2)^2}~\dot\rho,~~
g^a_{ij}=0~,~~g^a_{i4}=-g^a_{4i}~,~~i, j=1,2,3~.
$$
The adiabatic constraint $g^a_{\mu\nu}\ll G^a_{\mu\nu}$
means that the variation of instanton size is much
smaller than the characteristic transformation scale of the PP field, 
$\dot\rho\ll O(1).$ Then calculating the corrections for the 
action, it is reasonable to take out $\dot\rho$ beyond the integral and
the one instanton contribution to the action turns out to be
\begin{equation}
\label{8}
s_c=\frac14 \int d^4 x~ G'^2_{\mu\nu}\simeq\frac{8\pi^2}{g^2}+C~\dot\rho+
\frac{\kappa_{s.t.}}{2}~\dot\rho^2 ,
\end{equation}
where $\dot\rho$ should be taken as the mean rate of slow solution
deformation at a characteristic instanton lifetime $\sim \rho$. For
simplicity, one may take it in the centre of the instanton 
$\dot\rho (z)=\dot\rho (x,z)|_{x=z}.$ 
The constant $C=0$
(because the first variation of the action $\delta S/\delta A$
for the solution itself equals to zero).
For the 'kinematical' 
$\kappa$-term we have 
$\kappa_{s.t.}=
\frac{12\pi^2}{g^2}.$
The overt $\rho$ dependence of $\kappa$ is lacking because of the scale 
invariance. It arises with the renormalization of the coupling constant
(in a regular gauge the result is the same). 
Being within the ansatz (\ref{1}) we have
considered only 
the corrections induced by the variation of strength tensors, but not those
resulting from a possible variation of fields (\ref{2}). Bearing in mind
the form of potentials in regular ($r.g.)$ and singular $(s.g.)$ gauges
$$
A^{a}_\mu=\frac{1}{g} \eta_{a\mu\nu}~
\partial_\nu \ln(y^2+\rho^2)~~({\mbox{ r.g.}})~;
~~A^{a}_\mu=-\frac{1}{g} \bar \eta_{a\mu\nu}~
\partial_\nu \ln\left(1+\frac{\rho^2}{y^2}\right)~~({\mbox{ s.g.}})
$$
we find the adiabatic corrections $A_\mu^{'}=A_\mu + a_\mu$ as follows: 
\begin{equation}
\label{11}
a^{a}_\mu=\frac{2}{g} \eta_{a\mu 4}~
\frac{\rho}{y^2+\rho^2}~\dot\rho~~~~~~({\mbox{r.g.}})~.
\end{equation}
The substitution $\eta \to-\bar\eta$
brings about the transition from regular gauge to a singular one. 
Using the admixture $g_{\mu\nu}^a$ of chromoelectric and chromomagnetic
fields generated by the $a_\mu^{a}$ to saturate 
the functional integral as in Eq.(\ref{7}) we drop the terms of
higher order than $O(\dot\rho)$ out. Thus,
we come to the result for the 'kinematical' term  $\kappa$ as
{\footnote{In the singular gauge we have 
$\kappa=\frac32~\frac{32\pi^2}{g^2}.$}}
\begin{equation}
\label{12}
\kappa=\frac{32\pi^2}{g^2}~~~~~~({\mbox{r.g.}})~.
\end{equation} 
But within the accuracy taken at calculating Eq.(\ref{3}) we are 
permitted to fix $\kappa$ at some point as $\kappa(\bar\rho)$.

Analysing the variations of the functional integration result when
having calculated for the quasizero mode in the adiabatic approximation we
see that because of the kinetic energy smallness it is certainly permitted
to neglect its impact on the pre-exponential factor, which is small
itself. The inverse ~influence of the pre-exponential factor on the
'kinematical' term is negligible as well. Thus, going the way which 
we have already passed through while calculating Eq.(\ref{3}) we receive
within the required order of accuracy 
{\footnote{ 
For the case at hand the types of deformations (quasizero modes) may be 
simply counted in terms of the instanton solution. 
They are induced by the variations of the instanton size, the changes of
its position and colour 
orientation. In fact, there is one more type of the quasizero modes related 
to two far
distant instantons. 
}}
\begin{equation}
\label{13}
\langle S\rangle=\int d^4z \int d\rho~ n(\rho)~\left\{\frac12 \kappa~
\dot\rho^2+
s(\rho)\right\}~.
\end{equation}
It is important to remark here that the contribution of averaged PP 
pair interaction under the adiabatic conditions
is estimated as a term with contact interaction
\begin{equation}
\label{14}
\langle S(12) \rangle\simeq\int \frac{d^4 z}{V}
\frac{8\pi^2}{g^2}~\frac{\xi^2}{V}~\rho_1^{2}(z)~\rho_2^{2}(z)~.
\end{equation}
In particular, if the PP sizes do not vary 
we reproduce the well-known result \cite{2} 
$\langle S_{int}(12)\rangle=
\frac{8\pi^2}{g^2}~\frac{\xi^2}{V}~\rho_1^{2}\rho_2^{2}$
for such a contribution.

In the Minkowski space the factor in the curle brackets of Eq.(\ref{13}) 
might be interpreted as a mechanical system with Lagrangian
$
{\cal L}=\frac12~\kappa~\dot\rho^2 - U_{eff}(\rho)
$
if as a characteristic `velocity' 
$\partial \rho(x,z)/\partial x_4|_{x=z}$
we take a deformation of the field 
$\rho(x,z)|_{x=z}=\rho(z),
~\partial \rho/\partial x_4\sim\partial \rho/\partial z_4,$
and an action per one instanton might be taken as a 'potential energy'
$
U_{eff}(\rho)=\beta(\rho)+5\ln(\Lambda \rho)-\ln \widetilde \beta^{2N_c}
+\nu~\frac{\rho^2}{\overline{\rho^2}}~.
$
In the local vicinity of the potential minimum
$\rho_c^{2}=\frac{b-5}{2\nu}~\overline {\rho^2 },~
\left(\frac{U_{eff}}{d\rho}=0\right)$ the system is oscillating and
we have for the frequency (using the configuration corresponding
to Eq.(\ref{12}))
\begin{equation}
\label{15}
m^2=\frac{4\nu}{\kappa~\overline{\rho^2}}=\frac{\nu}{\beta~
\overline{\rho^2}}
\end{equation}
while calculating the second derivative
$\frac {d^2 U_{eff}(\rho)}{d\rho^2}\left.\right |_{\rho_c}=
\frac{4\nu}{\overline{\rho^2}}$.

We have only analysed the deformations in the temporary direction.
Those in the spatial directions could be estimated by drawing the same
arguments. Thus, the expression for the $\kappa$-term keeps 
the form obtained above with the only change of rates for the
appropriate gradients of function $\rho (x,z)$, i.e. the substitution 
$\dot \rho(t) \to \frac{\partial\rho (x,z)|_{x=z}}{\partial x}
\sim\frac{\partial\rho}{\partial z}$ should be 
performed for such a 'crumpled' instanton. 
Then the frequency of proper fluctuations might be interpreted 
as the mass term and the excitations occur to have a phonon-like nature
\begin{equation}
\label{16}
{\cal L}=
\frac12~\kappa~[\dot\rho^2-\nabla \rho\nabla \rho]-U_{eff}(\rho)~,
\end{equation}
(the cross-terms $\sim \dot\rho~\rho^{'}$ equal to zero identically)
{\footnote{It is interesting to remark that the centre of instanton solution 
may not be shifted since the relevant
deformations lead to the singular $\kappa$ (unlike the dilatational mode),
whereas the colour coordinate 
variation $\Omega$ gives the trivial result $\kappa=0$.}}.
With the instanton liquid parameters obtained above 
we have for the mass term  
$m\approx 1.21 \Lambda$. 
Then the wave length $\lambda_4\sim m^{-1}$ in the $x_4$-direction 
is $\lambda_4\Lambda\approx 0.83 \ge \bar V^{1/4}\Lambda  \approx 0.81 >
\bar\rho\Lambda$ (the size of the phonon localization in the spatial 
directions can be
arbitrary and may noticeably exceed $\lambda_4$, besides the number $N$ of 
the PP forming the excitations might be pretty large).   

These numerical values obtained should be taken rather qualitatively
illustrating the principle possibility to have the particle-like
excitations originated by the quasizero modes. It is clear that
striving
to go beyond the superposition ansatz one should to take into account at
least a medium change of the instanton profile and to develop
more realistic  description of the instanton interactions. Certainly, 
what presented here essentially exceeds the corresponding results which 
one could expect in the 'complete theory'.

In conclusion let us emphasize that we have considered the
excitations of the instanton liquid generated by the dilatational 
instanton deformations and the adiabatic assumption leads, in 
principle, to a fully consistent picture.
 The quantum numbers of the excitations are the scalar 'glueball' one 
and model itself regulates the most suitable regime of such 
phonon-like deformations resulting in mass gap generation fixed by
$\Lambda_{QCD}$.
This value of the mass scale looks very reasonable and absolutely 
self-consistent for an adiabatic approximation within the instanton picture 
of quenched QCD vacuum. Indeed, it is qualitatively clear that 'hard'
glueball mass $\sim 1.5~GeV$ (which is just known from existing
theoretical investigations in lattice QCD and sum rule methods) 
certainly leads beyond our imposed constraints. It means that
the strength of a deformation field becomes quite compatible with the 
instanton field and suppresses the $PP$.
But the 'hard' glueball scale is extracted from the exponential fall of
gluon correlation functions, and we believe that 
there is not well grounded to attribute this measurements a dynamical 
meaning of a signal propagation.

 Apparently, including the quark condensate
an intriguing guess is to associate some light hadrons 
with these phonon-like excitations discovered since the preliminary
evaluations of their mass spectrum look quite encouraging. 
We believe that we could imagine them as observable in mixing with the quark 
condensat, for example, with 'sigma' meson.
Moreover, the concept of the confining potential for the light quarks
in the context of our approach seems simply irrelevant because of a stable  
phonon nature.

The financial support of RFFI Grants 96-02-16303, 96-02-00088 G, 
97-02-17491 and INTAS Grants 93-0283, 96-0678 is greatly acknowledged.

\end{document}